\documentclass[%
  reprint,
  superscriptaddress,
  amsmath,
  amssymb,
  aps
]{revtex4-2}

\usepackage{graphicx}% Include figure files
\usepackage{dcolumn}% Align table columns on decimal point
\usepackage{bm}% bold math
\usepackage[colorlinks = True, linkcolor = red, citecolor = red]{hyperref}
\usepackage{graphicx}
\usepackage[normalem]{ulem}
\usepackage{xcolor}
\usepackage{enumitem}
\usepackage{orcidlink}
\usepackage[percent]{overpic}
\begin{document}

\title{Steady-State Spectral Kissing and Dissipative Phase Transition}

\author{Devesh Karthik\, 
\orcidlink{0009-0002-0565-3353}} 
\affiliation{Department of Physics, University of Connecticut, Storrs, Connecticut, USA}

\author{Jorge Ch\'avez-Carlos\,
\orcidlink{0000-0002-5223-5931}}
\affiliation{Physics Department, Cinvestav, AP 14-740, M\'exico City 07000, Mexico}

\author{Edson M. Signor\, \orcidlink{0009-0005-0011-415X}} 
\affiliation{Department of Physics, University of Connecticut, Storrs, Connecticut, USA}

\author{Francisco P\'erez-Bernal\,\orcidlink{0000-0002-3009-3696}}
\affiliation{Depto. de Ciencias Integradas y Centro de Estudios Avanzados en F\'isica, Matem\'aticas y Computaci\'on, Unidad Asociada GIFMAN CSIC-UHU, Universidad de Huelva, Huelva 21071, Spain}
\affiliation{Instituto Carlos I de F\'isica Te\'orica y Computacional, Universidad de Granada, Granada 18071, Spain}

\author{Victor S. Batista\,
\orcidlink{0000-0002-3262-1237}}
\affiliation{Department of Chemistry, Yale University, New Haven, Connecticut 06520, United States}
\affiliation{Yale Quantum Institute, Yale University, New Haven, Connecticut 06511, United States}

\author{Lea F. Santos\,\orcidlink{0000-0001-9400-2709}}  
%\email{lea.santos@uconn.edu} 
\affiliation{Department of Physics, University of Connecticut, Storrs, Connecticut, USA}

\begin{abstract}
Spectral kissing, recently probed in a Kerr parametric oscillator (KPO), refers to the merging of pairs of energy levels. It is a manifestation of excited-state quantum phase transitions (ESQPTs) associated with double-well phase-space structures. Here, we show that this phenomenon has a dissipative counterpart encoded in the spectrum of the steady-state density matrix. Using the dissipative KPO as a representative example, we demonstrate that, in the weak-dissipation regime, the eigenvalues of this matrix organize into quasi-degenerate pairs. By analyzing the classical limit, we show that this pairing arises from the two stable attractors that replace the minima of the isolated system, and that increasing dissipation progressively destroys the pairing as the stable attractors approach the origin. We estimate the number of eigenvalues exhibiting steady-state spectral kissing and analytically determine the critical dissipation strength at which the attractors coalesce with the origin. We further provide evidence that this coalescence is associated with a continuous dissipative phase transition. Our results indicate that signatures of ESQPTs associated with double- or multi-well phase-space structures can persist in open quantum systems and potentially be detected through steady-state tomography. 
\end{abstract}

\maketitle

\section{Introduction}
Spectral kissing is a term introduced in an experiment with a superconducting Kerr parametric oscillator (KPO) and refers to the exponential coalescence of pairs of energy levels~\cite{Frattini2024}. As shown in Ref.~\cite{Chavez2023}, this phenomenon is a manifestation of an excited-state quantum phase transition (ESQPT)~\cite{Cejnar2021}. While quantum phase transitions are identified through abrupt changes in the properties of the system's ground state as a control parameter is varied, ESQPTs extend this notion to excited states and depend on both the values of the energy and the control parameter.  For a fixed value of a control parameter, an ESQPT occurs at a critical energy value, $E_c$, where either the density of states or its derivative at a given order  exhibits a singularity associated with the clustering of energy levels \cite{Caprio2008, Cejnar2008, Cejnar2021}.
In systems with a single degree of freedom, such as a KPO, the clustering of eigenvalues is accompanied by the ``kissing'' of pairs of energy levels from the ground state up to $E_c$.

The origin of this behavior can be understood from the analysis of the classical limit of the system, whose Hamiltonian develops a double-well phase-space structure with two minima separated by a hyperbolic stationary point. The energy difference between the minima and the hyperbolic point defines the ESQPT critical energy. At this energy, the classical trajectories undergo a qualitative change from being confined around a single minimum to encompassing both minima. As the driving strength (control parameter) increases~\footnote{Beyond a threshold, however, the static effective Hamiltonian breaks down and chaos develops~\cite{Chavez2025}.}, the barrier between the two wells grows, resulting in an increasing number of pairs of exponentially close eigenvalues. In the KPO, the lowest-energy pair is particularly important, because it forms the logical basis of the Kerr-cat qubit~\cite{Cochrane1999,Goto2016,Puri2017,Grimm2020}, where quantum information is encoded in superpositions of two macroscopically distinct coherent states, suppressing local errors. 

In addition to KPOs, other prominent systems with double-well structures associated with ESQPTs include the Lipkin-Meshkov-Glick (LMG) model~\cite{Vidal2004b, Vidal2004c, Heiss2005,  Santos2016, Gamito2022, Khalouf2023} introduced in nuclear physics~\cite{Lipkin1965a,Lipkin1965b,Lipkin1965c} and vibron models describing molecular dynamics~\cite{Caprio2008, Bernal2008, Larese2011, Larese2013, KRivera2020, Dutta2024}. Similar phase-space structures also arise in chemical reactions, where the reactive flux crosses the barrier separating the two wells of a double-well potential energy surface along the reaction coordinate~\cite{Cabral2024,Dutta2024}.

ESQPTs can occur in the absence of a QPT~\cite{Stransky2021,Corps2022,Chavez2024} and have been associated with a variety of structural and dynamical phenomena in quantum systems~\cite{Cejnar2021}. These include eigenstate localization and anomalously slow dynamics near the critical energy~\cite{SantosBernal2015,Bernal2016,Santos2016,Chavez2023}, exponential growth of out-of-time-ordered correlators due to the underlying hyperbolic point~\cite{Pilatowsky2020,Chavez2023}, and applications to quantum-enhanced sensing~\cite{Li2023,Vijaywargia2026}. ESQPTs have also been linked to accelerated quantum evolution~\cite{Lobez2016,Kloc2018}, enhanced decoherence~\cite{Relano2008,Fernandez2009}, isomerization processes~\cite{Khalouf2019,Rafik2025}, Schr\"odinger-cat-state formation~\cite{Corps2022}, and tunneling phenomena~\cite{Nader2021,Venkatraman2023,Prado2023,Prado2025}.

The present work investigates whether signatures of spectral kissing can survive in the presence of coupling to an environment. In open quantum systems described by Lindblad master equations, the long-time behavior is governed by the steady-state density matrix $\rho_{\mathrm{ss}}$, corresponding to the zero-eigenvalue mode of the Liouvillian. If remnants of the ESQPT persist in the dissipative regime, they should be encoded in the structure of $\rho_{\mathrm{ss}}$.

To address this question, we use the KPO as a representative example of a system exhibiting an ESQPT associated with a double-well phase-space structure. We analyze the spectrum of the steady-state Hamiltonian $H_{\mathrm{ss}}$,  defined through $\rho_{\mathrm{ss}} = e^{-H_{\mathrm{ss}}}$, as a function of both the control parameter and the dissipation strength. This approach is  motivated by studies of open systems~\cite{Prosen2013,Moudgalya2019,Richter2025, Mondal2025a}, where the spectral properties of $H_{\mathrm{ss}}$ are used to characterize the onset of quantum chaos and thermalization. It differs from approaches based on the complex spectrum of the Liouvillian itself~\cite{Rubio2022,Iachello2024}. Although we focus on the double-well case, our analysis can be extended to other dissipative systems whose ESQPTs originate from phase-space structures with multiple wells, such as those discussed in~\cite{Prado2025,Vijaywargia2026}.

We find that, in the weak-dissipation regime, the spectrum of $H_{\mathrm{ss}}$ organizes into nearly degenerate pairs of eigenvalues, mirroring the spectral-kissing phenomenon of the corresponding closed systems. By analogy, we refer to this effect as ``steady-state spectral kissing,'' and trace its origin to the classical dissipative dynamics. While dissipation replaces the conservative trajectories of the closed system with flows toward stable attractors, the phase space retains a two-region structure. The dynamics is now governed by two stable attractors, associated with the minima of the isolated system and a hyperbolic fixed point at the origin.

This underlying phase-space structure leaves an imprint on the spectrum of the steady-state density matrix. We derive an analytical estimate for the onset of steady-state spectral kissing and determine its dependence on both the control parameter and the dissipation strength. As dissipation increases, the two stable attractors approach each other and eventually merge into a single stable attractor at the origin. We analytically determine the critical dissipation strength at which this coalescence occurs. This classical bifurcation marks the disappearance of the two nonzero attractors and the loss of the phase-space structure underlying steady-state spectral kissing. 

We show that, as the classical limit is approached, the threshold for the disappearance of steady-state spectral kissing converges toward the classical bifurcation point. Furthermore, the steady-state photon number approaches a classical limit that vanishes continuously but becomes nonanalytic at the critical dissipation strength, providing evidence for a continuous dissipative phase transition. These results demonstrate a connection between steady-state spectral kissing and dissipative criticality, suggesting that the spectrum of the steady-state density matrix can serve as a complementary probe to conventional approaches based on steady-state observables and the Liouvillian spectrum~\cite{Kessler2012,Horstmann2013,Carmichael2015,Fitzpatrick2017,Minganti2018,Minganti2023,Beaulieu2025}.

The dissipative dynamics of parametrically driven nonlinear oscillators has been extensively investigated in earlier works by Dykman~\cite{Dykman2012} and collaborators. They analyzed the classical phase-space structure of these systems, including the bifurcations of symmetry-related stable attractors and noise-induced switching between them~\cite{Dykman1998}. Marthaler and Dykman subsequently developed a semiclassical theory of the stationary distribution over quantum quasienergy states and demonstrated that switching between the two stable oscillatory states is dominated by quantum activation rather than tunneling when the relaxation rate exceeds the tunneling rate~\cite{Marthaler2006}. Further studies examined quantum heating and its signatures in noise and response spectra~\cite{Dykman2011}, as well as critical fluctuations and switching near the parametric excitation threshold~\cite{Lin2015}. These works established important properties of the stationary dynamics of parametrically driven oscillators. Here, as explained above, we focus on how the underlying double-well phase-space structure is reflected in the eigenvalue spectrum of the full steady-state density matrix and its connection to the spectral kissing associated with ESQPTs.

The work is organized as follows. In Sec.~II, we review spectral kissing in the isolated KPO and its connection to ESQPTs. In Sec.~III, we introduce dissipation and investigate the eigenvalue spectrum of the steady-state density matrix, identifying the emergence of steady-state spectral kissing. In Sec.~IV, we analyze the classical dissipative dynamics and derive an expression for the line of the steady-state spectral kissing, which separates paired and unpaired eigenvalues. In Sec.~V, we investigate the disappearance of the steady-state spectral kissing as dissipation increases and discuss its connection to the dissipative phase transition. Finally, Sec.~VI summarizes our conclusions.

\section{Isolated KPO: Spectral kissing} 

The experiment in which spectral kissing was probed consisted of a superconducting nonlinear asymmetric inductive element (SNAIL) transmon driven close to twice the natural frequency of the nonlinear oscillator~\cite{Frattini2024}. In the rotating frame, the system is described by an effective time-independent Hamiltonian. In units where $\hbar=1$, the  quantum Hamiltonian is 
\begin{equation}
\hat{H}
= K
\hat{a}^{\dagger 2}\hat{a}^2
-
\epsilon_2\left(\hat{a}^{\dagger 2}+\hat{a}^{2}\right),
\label{Eq:HamEff}
\end{equation}
where $\hat a^\dagger$ and $\hat a$ are bosonic creation and annihilation operators, $K$ is the Kerr nonlinearity, and  $\epsilon_2$ is the amplitude of the two-photon drive. We also define the dimensionless squeezing amplitude 
\begin{equation}
  \xi=\epsilon_2/K . 
\end{equation} The Hamiltonian conserves parity,
\[
[\hat H,(-1)^{\hat a^\dagger\hat a}]=0,
\]
so its eigenstates can be separated into even- and odd-parity sectors.

\begin{figure}[h]
    \centering
    \includegraphics[scale=0.34]{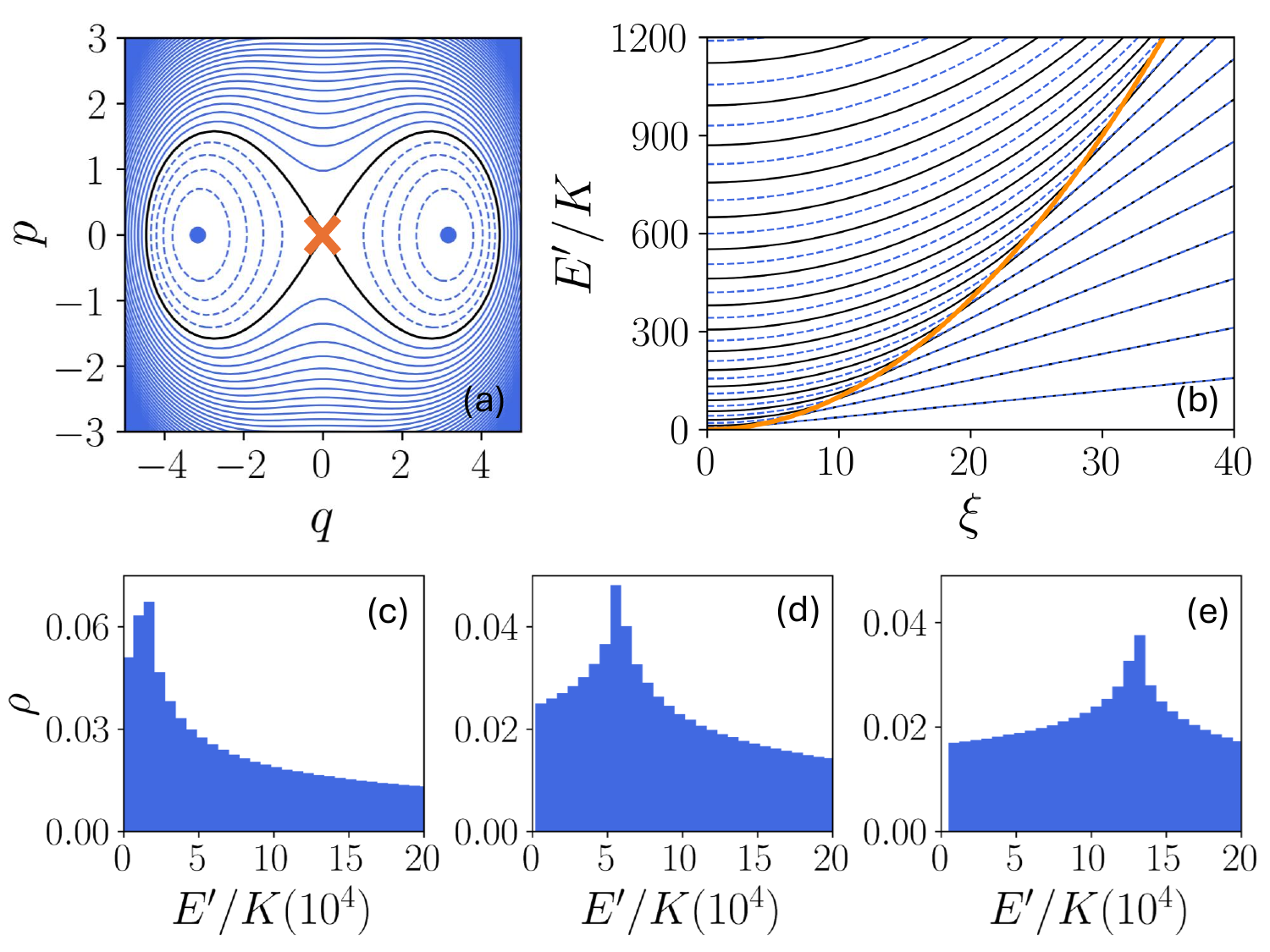}
    \caption{Isolated KPO. (a) Classical phase-space structure of the Hamiltonian in Eq.~\eqref{eq:hcl} for $\xi=5$. The blue dots mark the global minima, the orange cross at $(0,0)$ indicates the hyperbolic point, and the black curve crossing at the origin is the separatrix. (b) Excitation energies $E'=E-E_{\rm GS}$ of the quantum Hamiltonian in Eq.~\eqref{Eq:HamEff} as a function of the squeezing amplitude $\xi$, showing spectral kissing between the eigenvalues corresponding to even (solid lines) and odd (dotted lines) eigenstates. The orange line indicates the ESQPT critical energy $E'_{c}=K \xi^2$. (c)--(e) Density of states for (c) $\xi=120$, (d) $\xi=240$, (e) $\xi=360$, indicating the clustering of energy levels characteristic of ESQPTs. 
    \label{fig:Fig01}}
\end{figure}

The classical limit of Eq.~\eqref{Eq:HamEff}, derived in App.~\ref{AppClassH}, is obtained in the large-$\xi$ limit. Written in terms of the canonical phase-space variables $q$ and $p$, the classical Hamiltonian is given by
\begin{equation}
H_{\rm cl}
=
\frac{K}{4}(q^2+p^2)^2
-
\epsilon_2(q^2-p^2) .
\label{eq:hcl}
\end{equation}
It leads to a double-well phase-space structure for $\xi\neq 0$, as shown in Fig.~\ref{fig:Fig01}(a). The two wells are centered at the stable points $(q, \, p)=(\pm\sqrt{2\xi}, \, 0)$, while the origin is a hyperbolic stationary point, indicated by the orange cross. The black curve passing through the origin is the separatrix. Below the separatrix energy, the classical trajectories are confined to one of the two wells, forming pairs with the same energy. Above the separatrix, the trajectories encircle both wells.

The quantum counterpart of this phase-space structure is shown in Fig.~\ref{fig:Fig01}(b), which displays with solid  and dotted lines the excitation energies $E' = E-E_{\rm GS}$ 
of the quantum Hamiltonian in Eq.~\eqref{Eq:HamEff} as a function of $\xi$, where $E_{\rm GS}$ is the ground-state energy. Solid and dotted lines correspond to the even- and odd-parity sectors, respectively. The eigenvalues are obtained by diagonalizing the Hamiltonian in a truncated Fock basis, retaining only those that converged upon increasing the basis dimension.

The two classically disconnected regions below the separatrix in Fig.~\ref{fig:Fig01}(a) are reflected quantum mechanically in Fig.~\ref{fig:Fig01}(b) in pairs of nearly degenerate energy levels, giving rise to the phenomenon known as spectral kissing. As the squeezing amplitude $\xi$ increases, the wells become deeper and an increasing number of pairs exhibit this behavior. Spectral kissing extends up to the ESQPT critical excitation energy, 
\begin{equation}
E'_{c}=K\xi^2 ,    
\end{equation}
which corresponds classically to the energy difference between the separatrix and either of the two minima (see App.~\ref{AppB}).

A main signature of the ESQPT is the clustering of the eigenvalues around the critical energy, which produces a peak in the density of states, as shown in Figs.~\ref{fig:Fig01}(c)-(e) for different values of $\xi$. As the system approaches the classical limit, this peak sharpens and develops a logarithmic divergence.

The number of quantum levels below the separatrix can be estimated semiclassically using the Bohr--Sommerfeld quantization rule,
\[
\oint p\,dq \simeq 2\pi n.
\]
For a closed classical trajectory, the action $\oint p\,dq$ corresponds to the phase-space area enclosed by the trajectory. Thus, if $A$ denotes the total phase-space area enclosed by the separatrix, the number of individual energy eigenvalues exhibiting spectral kissing (those below the separatrix) can be estimated as $N_{\text{kiss}}^{(0)}\simeq A/(2\pi)$.

For the classical KPO Hamiltonian in Eq.~\eqref{eq:hcl}, the separatrix forms a Bernoulli lemniscate whose two lobes enclose a total area $A=4\xi$ (see App.~\ref{AppBernoulli}). It then follows that the number of quantum levels below $E'_c$ is given by 
\begin{equation}
N_{\text{kiss}}^{(0)}=\frac{2\xi}{\pi},
\label{Eq:nclosed}
\end{equation}
which increases linearly with the squeezing amplitude $\xi$.

\section{Open KPO: Steady-state spectral kissing} 

Following the experiment in Ref.~\cite{Albornoz2026}, we model the coupling of the KPO to the environment using a phenomenological Lindblad master equation,
\begin{equation}
    \frac{d\hat\rho}{dt}
    =
    \mathcal{L}\hat\rho
    =
    -i[\hat H,\hat\rho]
    + \kappa (1 + n_{\rm th}) \mathcal{D}[\hat{a}] \hat{\rho} + \kappa  n_{\rm th} \mathcal{D}[\hat{a}^\dagger]
    \hat{\rho},
    \label{Eq:Lindblad}
\end{equation}
where $\hat\rho$ is the density matrix, $\mathcal{L}$ is the
Liouvillian superoperator, and $\hat H$ is the Hamiltonian in Eq.~(\ref{Eq:HamEff}). The Lindblad dissipator associated with the operator
$\hat O$ is defined as
\[
\mathcal{D}[\hat O]\hat\rho
    =
    \hat O \hat\rho \hat O^\dagger
    - 
    (
        \hat O^\dagger \hat O \hat\rho
        +
        \hat\rho \hat O^\dagger \hat O
    )/2.
\]    
The parameter $\kappa$ is the single-photon dissipation rate, $n_{\rm th}$ is the effective thermal occupation, $\mathcal{D}[\hat a]$ describes single-photon loss, and $\mathcal{D}[\hat a^\dagger]$ indicates photon gain. As emphasized in Ref.~\cite{Albornoz2026}, this description is phenomenological and should not be interpreted as a microscopic thermalization model for the driven nonlinear KPO. A microscopic treatment of the system-environment coupling may become necessary beyond the weak-coupling regime or near quantum critical points~\cite{Kopylov2019,Wachtler2020,AntoSztrikacs2023}. We note, however, that a Lindblad formalism was also employed in the experimental investigation of dissipative phase transitions reported in Ref.~\cite{Beaulieu2025}.

We focus on the long-time state of the dissipative dynamics. The steady-state density matrix $\rho_{\rm ss}$  is determined by the kernel of the Liouvillian,
\begin{equation}
{\cal L}\rho_{\rm ss}=0,
\end{equation}
subject to the normalization condition $\mathrm{Tr}(\rho_{\rm ss})=1$. Analytical methods for obtaining $\rho_{\rm ss}$ have been developed for KPOs with one- and two-photon dissipation~\cite{Minganti2016,Roberts2020,Iachello2024}. In this work, we determine the steady-state density matrix numerically by solving the corresponding linear system in a truncated Fock basis. 

To investigate whether the pairwise spectral structure of the isolated KPO leaves a signature in the steady state, we consider the steady-state Hamiltonian $H_{\rm ss}$, defined through
\begin{equation}
    \rho_{\rm ss}=e^{-H_{\rm ss}}.
\end{equation}
Thus, if $p_j$ are the eigenvalues of $\rho_{\rm ss}$, the corresponding eigenvalues of $H_{\rm ss}$ are $\eta_j=-\ln p_j$. We use the shifted eigenvalues $\eta'_j=\eta_j-\eta_0$, where $\eta_0$ is the lowest eigenvalue of $H_{\rm ss}$. Convergence is checked by increasing the basis dimension, and only converged eigenvalues are included in the spectral analysis.

Figure~\ref{fig:Fig02}(b) shows $\eta'_j$ as a function of the squeezing amplitude $\xi$ for $\kappa/K=2$. Similarly to the energy spectrum of the isolated KPO in Fig.~\ref{fig:Fig01}(b), the eigenvalues organize into nearly degenerate pairs, with an increasing number of pairs appearing as $\xi$ increases. We refer to this pairwise organization of the steady-state spectrum as \emph{steady-state spectral kissing}. The eigenvalues also accumulate near the boundary separating the paired and unpaired regions of the spectrum, reminiscent of the level clustering associated with the ESQPT in the isolated system.

The similarity between Figs.~\ref{fig:Fig01}(b) and \ref{fig:Fig02}(b) raises the following question. What is the origin of the pairwise structure in the presence of dissipation, where the conservative trajectories and separatrix of the isolated KPO no longer characterize the dynamics? As we show in the next section, the answer is provided by the classical dissipative dynamics. Although dissipation replaces the two minima of the isolated system by stable attractors, the phase space retains a two-region structure that leaves a direct imprint on the spectrum of $H_{\rm ss}$.

\section{Open KPO: Classical limit} 

Insight into the origin of steady-state spectral kissing is provided by the classical limit of the open KPO, whose dynamics is governed by the dissipative flow (see App.~\ref{app:open_kpo_classical})
\begin{align}
\dot{q}
&=
2\epsilon_2 p
+
Kp(q^2+p^2)
-
\frac{\kappa}{2}q,
\\
\dot{p}
&=
2\epsilon_2 q
-
Kq(q^2+p^2)
-
\frac{\kappa}{2}p .
\end{align}
The flow has a constant negative divergence,
$\nabla\cdot(\dot q,\dot p)=-\kappa$,
reflecting the dissipative contraction of phase-space areas. The stationary points are 
\begin{eqnarray}
    x_1&=&(q_1,p_1)=\left(\sqrt{\frac{\sqrt f(\kappa_c+\sqrt f)}
{4K\kappa_c}},\sqrt{\frac{\sqrt f(\kappa_c-\sqrt f)}
{4K\kappa_c}},\right) ,\nonumber \\
    x_2&=&(-q_1,-p_1) ,\nonumber\\
    x_3&=&(0,0) ,
    \label{Eq:threepoints}
\end{eqnarray}
where
\begin{equation}
f=\kappa_c^2-\kappa^2,
\end{equation}
and
\begin{equation}   \kappa_c=4\epsilon_2
    \label{Eq:kappac}
\end{equation}
is a critical value.
The points $x_{1}$ and $x_{2}$ exist only for $\kappa<\kappa_c$. In this regime, $x_3$ is a hyperbolic fixed point, while $x_{1,2}$ are stable attractors. Specifically, trajectories approach these attractors through damped oscillations (stable spirals) for $\kappa < 2\kappa_c/\sqrt{5}$ or monotonically (stable nodes) for $2\kappa_c/\sqrt{5}< \kappa < \kappa_c$ (see App.~\ref{app:open_kpo_classical}).

\begin{figure}[t]
\includegraphics[scale=0.25]{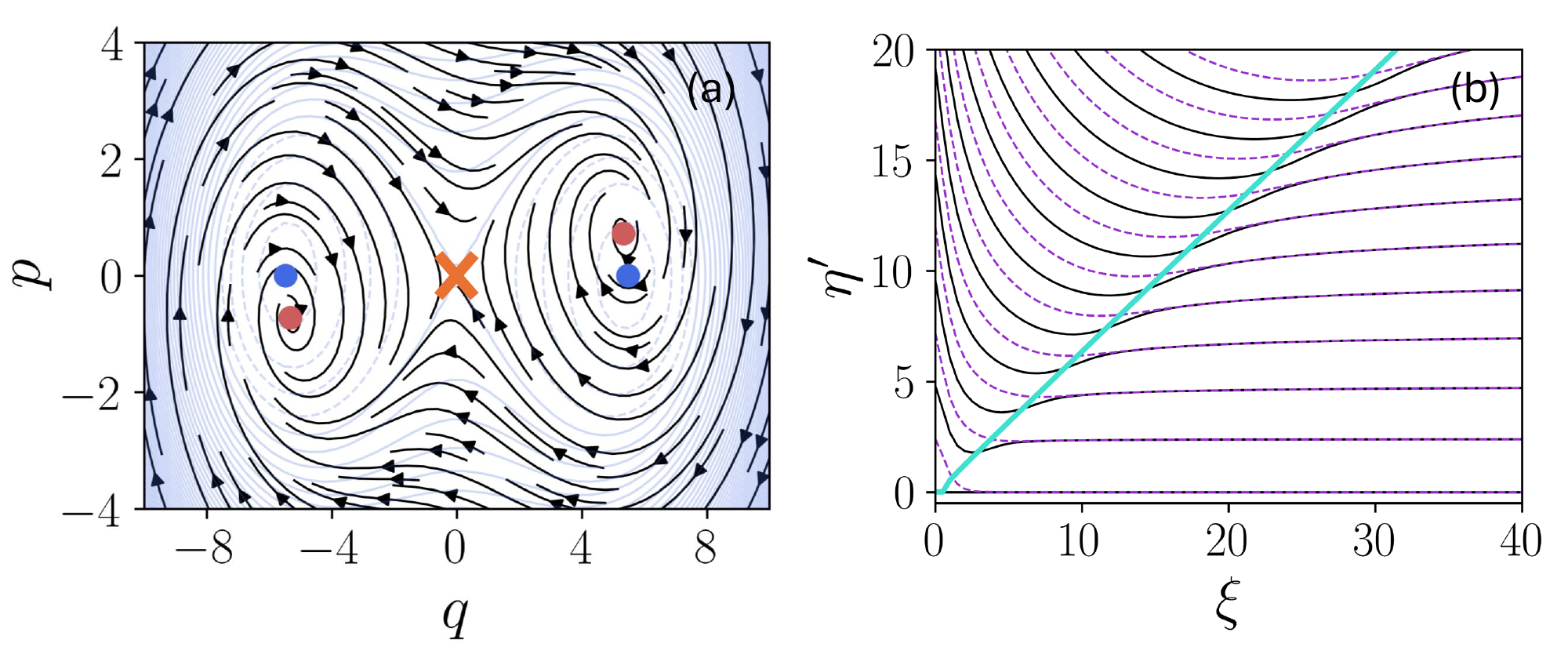}
    \caption{Open KPO for the parameters $\kappa/K=2$ and $n_{\rm th}=0.1$. (a) Classical phase-space dissipative flow  for $\xi=15$. The cross marks the hyperbolic point at the origin, red circles denote the stable spiral attractors, and blue circles indicate the minima of the closed system. (b) Eigenvalues $\eta'=\eta-\eta_0$ of the steady-state Hamiltonian $H_{\rm ss}$ vs. $\xi$, showing steady-state spectral kissing.  
    The blue line indicates the number $N_{\text{kiss}}^{(\text{ss})}$ of  eigenvalues of $H_{\rm ss}$ that form quasi-degenerate pairs, as defined in Eq.~(\ref{Eq_nss}).
    }    \label{fig:Fig02}
\end{figure}

Figure~\ref{fig:Fig02}(a) shows the phase space flow for $\kappa/K = 2 $ and $\xi=15$, where the red circles denote the stable spiral attractors, the blue circles indicate the minima of the closed system ($\kappa/K = 0$), and the hyperbolic point, marked with a cross, is common to both systems. The correspondence between the attractors of the dissipative system and the minima of the closed dynamics shows that, for weak dissipation, the phase space remains organized around two stable regions. This remnant of the double-well structure is responsible for the emergence of steady-state spectral kissing, as it induces pairs of quasi-degenerate eigenvalues of $H_{\rm ss}$.

The distance between an attractor and the origin is  
\begin{equation}
d=\sqrt{q_1^2+p_1^2}
=
\frac{f^{1/4}}{\sqrt{2K}}.
\label{Eq:d}
\end{equation}
In the limit $\kappa \rightarrow 0$, we see that the attractors approach the minima of the isolated KPO, located at $(q,p)=(\pm\sqrt{2\xi},0)$ and $d^2 = 2 \xi$, which coincides with the area $A/2$ enclosed by a single lobe of the separatrix. This correspondence suggests using $d^2$ as the natural measure of the effective phase-space region associated with each attractor in the dissipative system. By analogy with the semiclassical counting of energy levels for the closed KPO [Eq.~(\ref{Eq:nclosed})], we therefore introduce 
\begin{equation}
    N_{\text{kiss}}^{(\text{ss})} = \frac{2 d^2}{2 \pi} = \frac{\sqrt{ f }}{2 \pi K} = \frac{2\xi}{\pi}
\sqrt{1-\left(\frac{\kappa}{\kappa_c}\right)^2} ,
\label{Eq_nss}
\end{equation}
which provides an estimate for the number of individual eigenvalues of $H_{ss}$ that form quasi-degenerate pairs due to the underlying two-attractor phase-space structure. 

The expression in Eq.~(\ref{Eq_nss}) is shown as the blue line in Fig.~\ref{fig:Fig02}(b). This lines provides an estimate
for the boundary between the paired and unpaired
regions of the spectrum of $H_{\rm ss}$, marking the point of the steady-state spectral kissing. As $\xi$ increases, the attractors move farther apart, enlarging the effective phase-space area associated with the two-attractor structure and increasing the number of eigenvalues that
form quasi-degenerate pairs.

For large $\xi$, the neighboring pairs of quasi-degenerate eigenvalues in Fig.~\ref{fig:Fig02}(b) become approximately equally spaced, with separation $\ln[(n_{\rm th}+1)/n_{\rm th}]$, as derived in App.~\ref{Appendix:harmonic-oscillator} for the weak-dissipation regime. This is consistent with the description of stationary quasienergy distributions near stable states of parametrically driven nonlinear oscillators~\cite{Marthaler2006}, and indicates that, in the large-$\xi$ limit, the local dynamics around each of the two attractors is approximately harmonic.

\section{Attractor coalescence and dissipative phase transition} 

Contrary to the isolated KPO, where the double-well structure exists for any $\xi \neq 0$ and the ground state is twofold degenerate, the emergence of steady-state spectral kissing requires the two-photon drive to overcome the stabilizing effect of one-photon dissipation. The two-photon drive destabilizes the origin and generates two symmetry-related stable attractors, whereas dissipation tends to restore the stability of the origin. The existence of the two attractors, which provide the classical counterpart of the nearly degenerate steady-state spectral pairs, is therefore determined by the competition between driving and dissipation.

This competition can be understood from the classical fixed points. According to Eqs.~(\ref{Eq:threepoints})–(\ref{Eq:kappac}) and Eq.~(\ref{Eq:d}), the two nonzero stable attractors exist only when the distance $d \neq 0$, that is,
\begin{equation}
    \frac{\kappa}{K}<\frac{\kappa_c}{K} = 4\xi .
\end{equation}
Increasing the dissipation strength has two simultaneous effects: the attractors move toward the origin, as their radial distance decreases, and they rotate in phase space. To characterize this rotation, consider the attractor at $x_1=(q_1,p_1)$, with $q_1>0$, and define its angle $\vartheta_1$ relative to the positive $q$ axis. Its orientation is given by
\begin{equation}
\tan\vartheta_1
=\frac{p_1}{q_1}
=\frac{\kappa_c-\sqrt f}{\kappa}
=\frac{\kappa}{\kappa_c+\sqrt{\kappa_c^2-\kappa^2}}.
\end{equation}
The second attractor is located at $x_2=(-q_1,-p_1)$, as required by the symmetry of the system, and its angle is therefore $\vartheta_2=\vartheta_1+\pi$.

In the absence of dissipation, $\kappa=0$, the attractors lie on the $q$ axis, at a distance $d=\sqrt{2\xi}$ from the origin, with $\vartheta_1=0$ and $\vartheta_2=\pi$. As $\kappa$ increases, both attractors move inward and rotate counterclockwise. In the limit $\kappa\to\kappa_c^{-}$, their distance from the origin vanishes, while their angles approach $\vartheta_1=\pi/4$ and $\vartheta_2=5\pi/4$. Thus, the two attractors approach the origin from opposite directions along the diagonal $p=q$ and eventually coalesce with it at the critical dissipation strength, when the origin changes from a hyperbolic fixed point to a stable attractor. Beyond this threshold, the two nonzero attractors no longer exist, providing a semiclassical explanation for the disappearance of steady-state spectral kissing.

\begin{figure}[t]
    \centering
    \includegraphics[scale=0.35]{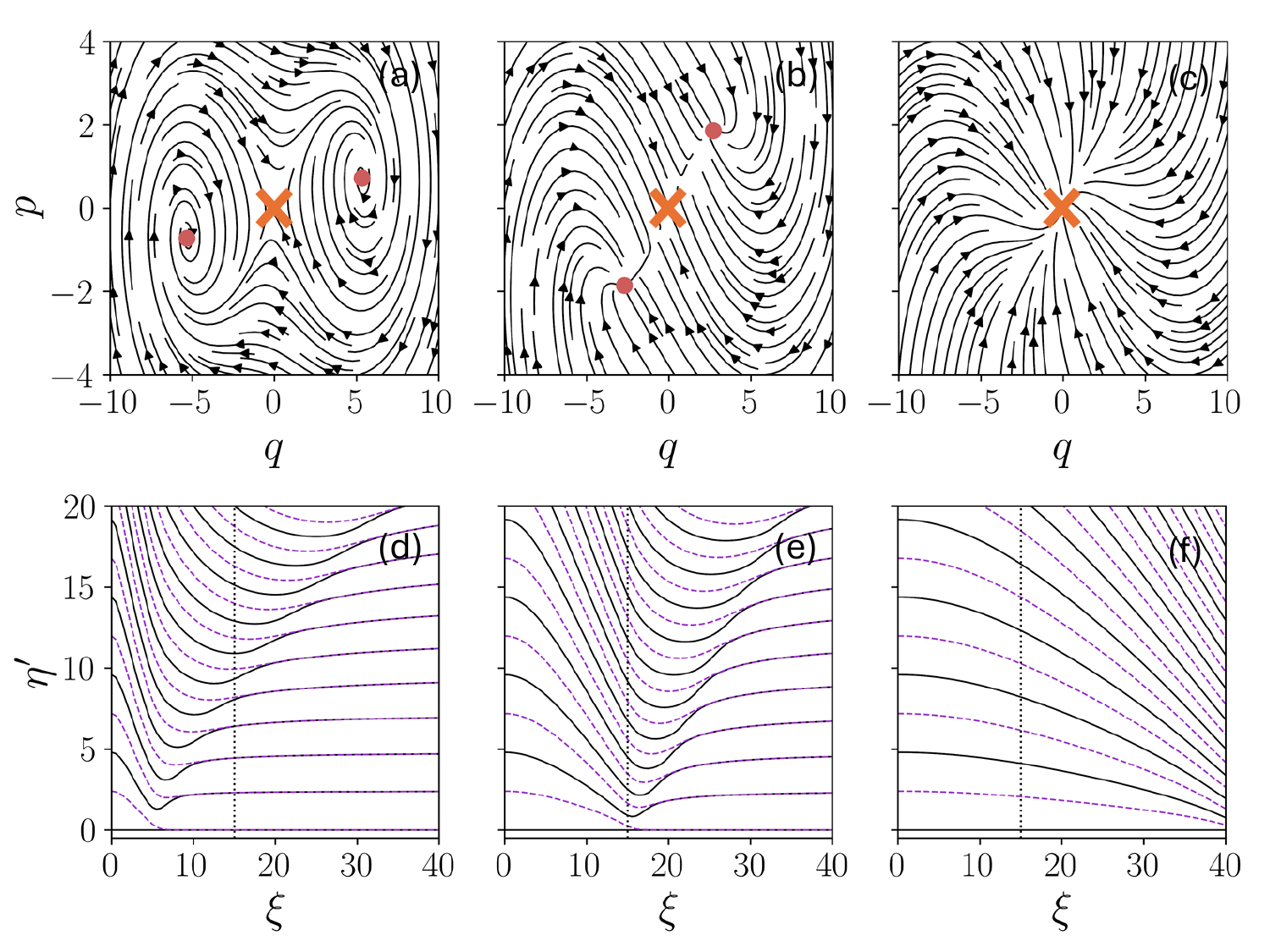}
    \caption{Open KPO for $n_{\rm th}=0.1$ and increasing dissipation strength; from left to right: $\kappa/K=16$, $56$, and $160$. (a)-(c) Classical phase space dissipative flow for $\xi=15$, corresponding to $\kappa_c/K=60$. The cross marks the fixed point at the origin, which is hyperbolic in (a)-(b) and becomes a stable attractor in (c). The red circles denote (a) stable spirals and (b) stable nodes. (d)-(f) Eigenvalues $\eta'=\eta-\eta_0$ of the steady-state Hamiltonian $H_{\rm ss}$ as a function of  $\xi$. The vertical line indicates the value of $\xi=15$ used in the corresponding upper panels (a)-(c).
   }
    \label{fig:Fig03}
\end{figure}

The merging process of the attractors $x_{1}$ and $x_{2}$, together with the change in their orientation,  is illustrated in Figs.~\ref{fig:Fig03}(a)-(c) for $\xi =15$, where the dissipation strength $\kappa/K$ increases from Fig.~\ref{fig:Fig03}(a) to Fig.~\ref{fig:Fig03}(c). The corresponding plots for the eigenvalues $\eta'$ as a function of $\xi$ are shown in Figs.~\ref{fig:Fig03}(d)-(f). As the attractors approach each other, the region of paired eigenvalues progressively shrinks. In Fig.~\ref{fig:Fig03}(f), where $\kappa/K \geq \kappa_c/K$ for all values of $\xi$ shown in the panel, the spectrum exhibits no steady-state spectral kissing, consistent with the absence of attractors away from the origin. 

\begin{figure}[h]
    \includegraphics[scale=0.3]{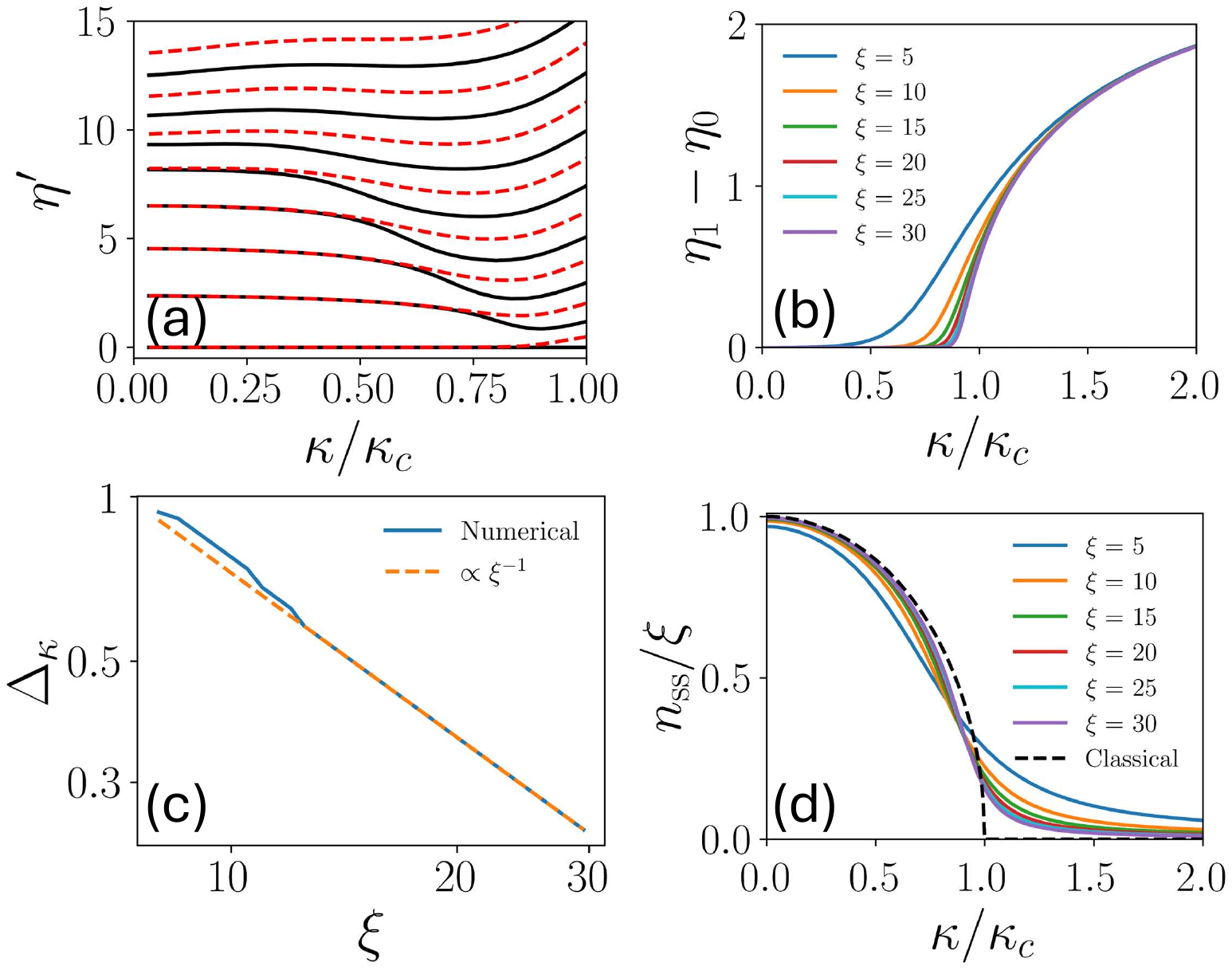}
    \caption{Open KPO for $n_{\rm th}=0.1$. (a) Shifted eigenvalues $\eta' = \eta -\eta_0$ of the steady-state Hamiltonian $H_{\rm ss}$ as a function of $\kappa / \kappa_c$ for $\xi=15$. (b) Splitting $\eta_1 -\eta_0$ between the two lowest eigenvalues of $H_{\rm ss}$ as a function of $\kappa / \kappa_c$ for increasing values of $\xi$. (c) Relative deviation $\Delta_{\kappa}=1-\kappa^*/\kappa_c$ from the classical critical point as a function of $\xi$, where $\kappa^*$ is the dissipation strength at which the splitting $\eta_1-\eta_0$ first exceeds a small threshold (see text). The power-law scaling $\Delta_{\kappa} \propto 1/\xi$ demonstrates that the threshold for the disappearance of steady-state spectral kissing converges toward the classical bifurcation point $\kappa=\kappa_c$ as $\xi\to\infty$. (d) Rescaled quantum steady-state photon number $n_{\rm ss}/\xi$ as a function of $\kappa/\kappa_c$ for increasing values of $\xi$, compared with the classical prediction $n_{\rm cl}/\xi$ from Eq.~(\ref{Eq:classical_photon}). The convergence toward the classical result provides evidence for a dissipative phase transition.
    }
    \label{fig:Fig04}
\end{figure}

To examine this process more closely, Fig.~\ref{fig:Fig04}(a) shows the eigenvalues of $H_{\rm ss}$ as a function of $\kappa/\kappa_c$ for fixed $\xi=15$. As dissipation increases, the quasi-degenerate pairs progressively split, beginning with the highest eigenvalues and extending toward the lowest ones. The lowest pair remains nearly degenerate until the dissipation strength approaches $\kappa_c$.

The behavior of the lowest pair is examined in Fig.~\ref{fig:Fig04}(b), which shows the splitting $\eta_1-\eta_0$ as a function of $\kappa/\kappa_c$ for increasing values of $\xi$. At finite $\xi$, the splitting becomes appreciable before the classical bifurcation point is reached. However, as $\xi$ increases and the system approaches the classical limit, the onset of the splitting shifts toward $\kappa/\kappa_c=1$, indicating that the disappearance of steady-state spectral kissing converges toward the classical bifurcation at $\kappa=\kappa_c$.

% \begin{figure}[t]
%     \centering
%     \begin{overpic}[width=0.52\columnwidth]{SSK_Fig_4a.pdf}
%         \put(25,25){\small (a)}
%     \end{overpic}\hfill
%     \begin{overpic}[width=0.48\columnwidth]{SSK_Fig_4b.pdf}
%         \put(21,27){\small (b)}
%     \end{overpic}

%     \vspace{2mm}

%     \begin{overpic}[width=0.48\columnwidth]{SSK_Fig_4c.pdf}
%         \put(27,25){\small (c)}
%     \end{overpic}\hfill
%     \begin{overpic}[width=0.50\columnwidth]{SSK_Fig_4d.pdf}
%         \put(25,25){\small (d)}
%     \end{overpic}

%     \caption{(a) ... (b) ... (c) ... (d) ...}
%     \label{fig:four_panels}
% \end{figure}

To quantify this behavior, we define $\kappa^*$ as the dissipation strength at which the splitting between the two lowest eigenvalues, $\eta_1-\eta_0$, first exceeds a threshold. The choice of this threshold depends on $\xi$ because the degree of near-degeneracy increases as the classical limit is approached. For larger $\xi$, the two eigenvalues remain extremely close over a broad range of dissipation strengths below $\kappa_c$, allowing the use of a smaller threshold. We verified that, for $\xi>15$, thresholds below $10^{-6}$ yield consistent values of $\kappa^*$ and do not affect the extracted scaling exponent. For smaller $\xi$, larger thresholds are needed to identify the disappearance of the near-degeneracy.

Figure~\ref{fig:Fig04}(c) shows the relative deviation 
\begin{equation}
\Delta_\kappa=1-\kappa^*/\kappa_c
\end{equation}
from the classical critical point as a function of $\xi$. We find that this deviation decreases algebraically with increasing $\xi$, following the scaling law 
\begin{equation}
\Delta_\kappa\propto\xi^{-\nu}, 
\end{equation}
where $\nu=1$ is the fitted exponent. This scaling demonstrates that the threshold for the disappearance of steady-state spectral kissing converges toward the classical bifurcation point $\kappa=\kappa_c$ as $\xi\to\infty$.

\subsection{Dissipative phase transition}

The disappearance of steady-state spectral kissing at the classical bifurcation suggests a connection with a dissipative phase transition. Dissipative phase transitions in two-photon-driven Kerr resonators were investigated experimentally~\cite{Beaulieu2025} through steady-state photon-number measurements and the analysis of the lowest eigenvalues of the Liouvillian. Here, our goal is to establish a connection between this critical behavior and the disappearance of quasi-degenerate pairs in the eigenvalue spectrum of the steady-state density matrix.

To investigate this connection, we examine the steady-state photon number,
\begin{equation}
n_{\rm ss}=\operatorname{Tr}(\rho_{\rm ss}a^\dagger a),
\end{equation}
which provides a physical observable that can be compared directly with the classical dynamics. Since the photon number grows with the squeezing parameter $\xi$, we consider the rescaled quantity $n_{\rm ss}/\xi$ and investigate its behavior as $\xi\to\infty$.

The classical photon number is related to the phase-space coordinates by
\begin{equation}
n_{\rm cl}=\frac{q^2+p^2}{2}.
\label{Eq:ncl}
\end{equation}
This relation follows from the definition of the bosonic annihilation operator, $\hat a=(\hat q+i\hat p)/\sqrt{2}$ (see App.~\ref{AppClassH}), which gives $\hat a^\dagger\hat a=(\hat q^2+\hat p^2-1)/2$. In the classical limit, the vacuum contribution becomes negligible, yielding Eq.~(\ref{Eq:ncl}). 

For $\kappa<\kappa_c$, the two stable attractors have the same distance $d$ from the origin, given by Eq.~(\ref{Eq:d}).  Since $q_1^2+p_1^2=q_2^2+p_2^2=d^2$, we obtain
\begin{equation}
n_{\rm cl}=\frac{d^2}{2}
=\frac{\sqrt{\kappa_c^2-\kappa^2}}{4K}.
\end{equation}
Using $\kappa_c=4K\xi$, and noting that the only stable attractor for $\kappa\geq\kappa_c$ is the origin, we obtain
\begin{equation}
\frac{n_{\rm cl}}{\xi}=
\begin{cases}
\displaystyle\sqrt{1-\left(\frac{\kappa}{\kappa_c}\right)^2},
& \kappa<\kappa_c,\\[6pt]
0,&\kappa\geq\kappa_c.
\end{cases}
\label{Eq:classical_photon}
\end{equation}

Figure~\ref{fig:Fig04}(d) shows the rescaled quantum steady-state photon number $n_{\rm ss}/\xi$ as a function of $\kappa/\kappa_c$ for increasing values of $\xi$, together with the classical prediction in Eq.~(\ref{Eq:classical_photon}). As $\xi$ increases, the quantum results converge toward the classical curve. For $\kappa<\kappa_c$, the rescaled photon number remains finite, reflecting the presence of the two nonzero stable attractors. For $\kappa>\kappa_c$, it approaches zero, consistent with the stabilization of the origin.

The rescaled classical photon number vanishes continuously at $\kappa=\kappa_c$, but its derivative with respect to $\kappa/\kappa_c$,
\begin{equation}
\frac{d}{d(\kappa/\kappa_c)}
\left(\frac{n_{\rm cl}}{\xi}\right)
=
-\frac{\kappa/\kappa_c}
{\sqrt{1-(\kappa/\kappa_c)^2}},
\qquad \kappa<\kappa_c,
\end{equation}
diverges as the critical point is approached from below, $\kappa\to\kappa_c^-$, revealing a nonanalyticity characteristic of a continuous phase transition.
Near the critical point,
\begin{equation}
\frac{n_{\rm cl}}{\xi}
\sim\sqrt{2\left(1-\frac{\kappa}{\kappa_c}\right)},
\end{equation}
corresponding to a critical exponent of $1/2$.

The convergence of the quantum steady-state photon number toward the classical result in Eq.~(\ref{Eq:classical_photon}), which becomes nonanalytic at $\kappa=\kappa_c$, provides evidence for a continuous dissipative phase transition in the classical limit. The rescaled photon number acts as an order-parameter-like quantity, remaining finite below the critical dissipation strength and vanishing continuously at $\kappa_c$. Together with the spectral results in Figs.~\ref{fig:Fig04}(a)--(c), this establishes a connection between the disappearance of steady-state spectral kissing and the dissipative phase transition associated with the coalescence of the classical attractors.

%%%%%%%%%%%%%%%%%%%%%%%%
\section{Conclusions}
We demonstrated that spectral kissing associated with excited-state quantum phase transitions (ESQPTs) has a dissipative counterpart encoded in the spectrum of the steady-state Hamiltonian $H_{\rm ss}$, defined through the steady-state density matrix $\rho_{\rm ss}=e^{-H_{\rm ss}}$. Just as the energy levels of the isolated system form nearly degenerate pairs, so do the eigenvalues of $H_{\rm ss}$. We showed that this pairing arises from a two-attractor phase-space structure inherited from the double-well landscape underlying the ESQPT. 

Using the classical limit, we obtained an analytical estimate for the number of eigenvalues exhibiting steady-state spectral kissing. As dissipation increases, the spectral pairing progressively disappears, reflecting the approach of the two stable attractors toward the origin. We analytically determined the critical dissipation strength at which the attractors coalesce.

We established a connection between the disappearance of steady-state spectral kissing and a continuous dissipative phase transition. First, we showed that, as the classical limit is approached, the threshold for the loss of spectral pairing converges toward the classical bifurcation point, where the attractors coalesce. Second, we showed that the rescaled steady-state photon number approaches a classical limit that becomes nonanalytic at the critical dissipation strength, vanishing continuously with a critical exponent of $1/2$. This convergence provides evidence for a continuous dissipative phase transition. These results demonstrate that the spectrum of the steady-state density matrix provides an alternative means of probing dissipative criticality, without requiring an analysis of the Liouvillian spectrum.

Our findings indicate that spectral signatures of ESQPTs associated with double- or multi-well phase-space structures can persist in dissipative quantum systems. The possibility of reconstructing the steady-state density matrix through quantum-state tomography~\cite{Kono2018} makes these signatures potentially accessible experimentally.

\section{Acknowledgments}

This work was supported by the Department of Energy grant number DE-SC0026657. JC-C has received a fellowship (No. 493338) from Secretaría de Ciencias y Humanidades, Tecnología e Innovación (SECIHTI). E.M.S. was supported by the
Research Corporation Cottrell SEED award CS-SEED-2025-003. F.P.-B. acknowledges funding through Grant No. PID2022-136228NB-C21 funded by
MICIU/AEI/10.13039/501100011033 and, as appropriate, by “ERDF A way of making Europe, by ERDF/EU,” by the European Union, or by the European Union NextGenerationEU/PRTR and by the Consejería de Conocimiento, Investigación y Universidad, Junta de Andalucía and European Regional Development Fund, through Grant No. UHU-1262561. Computing resources supporting this work were partly provided by the CEAFMC and Universidad de Huelva HPC cluster located in the Campus Universitario “El Carmen” and funded by FEDER/MINECO Project No. UNHU-15CE-2848. We thank Oscar Rosas-Ortiz for discussions regarding the distance between an attractor and the origin in the system considered. L.F.S. thanks Masudul Haque for discussions about the steady-state Hamiltonian. She is also grateful to the Mainz Institute for Theoretical Physics (MITP) of the Cluster of Excellence PRISMA+ (Project ID 390831469), for enabling her to complete a portion of this work.

%%%%%%%%%%%%%%%%%%%%%%%%%%%%%%
%%%%%%%%%%%%%%%%%%%%%%%%%%%%%%

\appendix

%%%%%%%%%%%%%%%%%%%%%%%%%%%%%%
%%%%%%%%%%%%%%%%%%%%%%%%%%%%%%

\section{Isolated Classical Hamiltonian}
\label{AppClassH}

We start from the quantum Hamiltonian
\begin{equation}
\frac{\hat H}{\hbar}
=
K\hat a^{\dagger 2}\hat a^2
-
K\xi\left(\hat a^{\dagger 2}+\hat a^2\right),
\label{EqSM_Hq}
\end{equation}
and work in units where $\hbar=1$.

To define the classical limit, we introduce the canonical position and momentum operators $\hat q$ and $\hat p$ through
\begin{equation}
\hat a
=
\frac{\hat q+i\hat p}{\sqrt{2\hbar_{\rm eff}}},
\qquad
\hat a^\dagger
=
\frac{\hat q-i\hat p}{\sqrt{2\hbar_{\rm eff}}},
\end{equation}
so that
\begin{equation}
[\hat q,\hat p]=i\hbar_{\rm eff}.
\end{equation}
Substituting the position and momentum operators into the Kerr term gives
\begin{align}
\hat a^{\dagger 2}\hat a^2
&=
\frac{1}{4\hbar_{\rm eff}^2}
(\hat q-i\hat p)^2(\hat q+i\hat p)^2 ,
\end{align}
and 
the two-photon driving term becomes
\begin{align}
\hat a^{\dagger 2}+\hat a^2
&=
\frac{1}{2\hbar_{\rm eff}}
\left[
(\hat q-i\hat p)^2
+
(\hat q+i\hat p)^2
\right]  ,
\end{align}
so the quantum Hamiltonian is written as
\begin{equation}
  \hat{H} =   \frac{K}{4\hbar_{\rm eff}^2}
(\hat q-i\hat p)^2(\hat q+i\hat p)^2 - \frac{K \xi}{2\hbar_{\rm eff}}
\left[
(\hat q-i\hat p)^2
+
(\hat q+i\hat p)^2 \right] .
\end{equation}

The classical limit is obtained by taking
\begin{equation}
\hbar_{\rm eff}\to 0,
\end{equation}
in which case $\hat q\to q$ and $\hat p\to p$ become classical phase-space variables.
This implies that
\begin{equation}
(\hat q-i\hat p)^2(\hat q+i\hat p)^2 
\longrightarrow
(q^2+p^2)^2
\end{equation}
and
\begin{align}
(\hat q-i\hat p)^2
+
(\hat q+i\hat p)^2
 \longrightarrow
2(q^2-p^2).
\end{align}
Furthermore, to obtain a finite classical Hamiltonian as
$\hbar_{\rm eff}\to 0$, we introduce the scaled parameters
\begin{equation}
K
=
K_{\rm cl} \, \hbar_{\rm eff}^2,
\qquad
\xi
=
\frac{\xi_{\rm cl}}{\hbar_{\rm eff}} .
\end{equation}
As a result, the classical Hamiltonian is
\begin{equation}
H_{\rm cl} =
\frac{K_{\rm cl}}{4}(q^2+p^2)^2
-
K_{\rm cl}\xi_{\rm cl}(q^2-p^2).
\label{Eq:SM_Hcl}
\end{equation}

The formal classical limit is obtained by taking $\hbar_{\rm eff}\to 0$ while keeping $K_{\rm cl}$ and $\xi_{\rm cl}$ fixed. Since $\xi=\xi_{\rm cl}/\hbar_{\rm eff}$, this limit corresponds to $\xi\to\infty$. Alternatively, the classical limit can be approached by 
\[
\boxed{\text{fixing } \hbar_{\rm eff}=1 \text{ and increasing } \xi}, 
\]
which is more directly relevant to experiments.

From now on, we omit the subscript ``cl'' and denote the parameters of the classical system by the same symbols used in the quantum model.

\section{Stationary points}
\label{AppB}

The classical Hamiltonian develops a double-well phase-space structure for $\xi \neq 0$. Using that $K\xi=\epsilon_2$, the stationary points are obtained
from
\begin{align}
\frac{\partial H_{\rm cl}}{\partial q}
&=
Kq(q^2+p^2)-2\epsilon_2 q=0,
\label{Eq:dHdq_stationary}\\
\frac{\partial H_{\rm cl}}{\partial p}
&=
Kp(q^2+p^2)+2\epsilon_2 p=0.
\label{Eq:dHdp_stationary}
\end{align}
These equations can be written as
\begin{align}
q\left[K(q^2+p^2)-2\epsilon_2\right]&=0,
\\
p\left[K(q^2+p^2)+2\epsilon_2\right]&=0.
\end{align}
Here, we assume that $K>0$ and $\epsilon_2>0$, so the second
factor in the second equation,
$K(q^2+p^2)+2\epsilon_2$, is always positive.
Therefore, the second equation requires
$p=0$.
Substituting $p=0$ into the first equation gives
\begin{equation}
q\left(Kq^2-2\epsilon_2\right)=0.
\end{equation}
Thus, there are three stationary points:
\begin{equation}
    (q,p)=(0,0)
\end{equation}
and
\begin{equation}
(q,p)=
\left(\pm\sqrt{\frac{2\epsilon_2}{K}},0\right)=\left(\pm\sqrt{2\xi},0\right).
\label{Eq:MinimaKPOisolated}
\end{equation}

To determine the character of these stationary points, we examine
the Hessian matrix of $H_{\rm cl}$,
\begin{equation}
\mathcal{H}
=
\begin{pmatrix}
\partial_q^2 H_{\rm cl} & \partial_q\partial_p H_{\rm cl} \\
\partial_p\partial_q H_{\rm cl} & \partial_p^2 H_{\rm cl}
\end{pmatrix},
\end{equation}
where
\begin{align}
\partial_q^2 H_{\rm cl}
&=
K(3q^2+p^2)-2\epsilon_2,
\\
\partial_p^2 H_{\rm cl}
&=
K(q^2+3p^2)+2\epsilon_2,
\\
\partial_q\partial_p H_{\rm cl}
&=
2Kqp.
\end{align}
At the origin, the Hessian is
\begin{equation}
\mathcal{H}(0,0)
=
\begin{pmatrix}
-2\epsilon_2 & 0 \\
0 & 2\epsilon_2
\end{pmatrix}.
\end{equation}
Its eigenvalues have opposite signs, indicating that the origin is the hyperbolic point separating the two wells.

At the two nonzero stationary points, the Hessian becomes
\begin{equation}
\mathcal{H}(q_0,0)
=
\begin{pmatrix}
4\epsilon_2 & 0 \\
0 & 4\epsilon_2
\end{pmatrix}.
\end{equation}
Both eigenvalues are positive, and therefore these two stationary
points are minima of the classical Hamiltonian. The isolated KPO
therefore has a double-well phase-space structure, with two minima located at the points given by Eq.~\eqref{Eq:MinimaKPOisolated} and a hyperbolic point
at the origin.

The energy at the hyperbolic point is
\begin{equation}
E_{\rm hyp}
=
H_{\rm cl}(0,0)
=
0,
\end{equation}
while the energy at either minimum is
\begin{equation}
E_{\rm min}
=
H_{\rm cl}(q_0,0)
=
-\frac{(\epsilon_2)^2}{K} .
\end{equation}
The energy difference between the hyperbolic point and the minima defines the ESQPT critical energy,
\begin{equation}
E_c'=
\frac{(\epsilon_2)^2}{K}
=
K \xi^2.
\label{Eq:Ec_classical}
\end{equation}

\section{Area below the separatrix}
\label{AppBernoulli}

Using polar coordinates in phase space,
\begin{equation}
q=r\cos\theta,
\qquad
p=r\sin\theta,
\label{EqSM_polar}
\end{equation}
the classical Hamiltonian in Eq.~(\ref{Eq:SM_Hcl})  becomes
\begin{align}
h_{\rm cl}
&=\frac{H_{\rm cl}}{K}=
\frac{r^4}{4}
-
\xi r^2 \cos(2\theta).
\end{align}

The separatrix corresponds to the critical energy of the hyperbolic point,
\begin{equation}
h_{\rm cl}=0.
\end{equation}
Substituting this condition into the classical Hamiltonian yields
\begin{equation}
\frac{r^4}{4}
-
\xi r^2 \cos(2\theta)
=0.
\end{equation}
Besides the trivial solution $r=0$, corresponding to the hyperbolic point itself, the separatrix is described by
\begin{equation}
r^2
=
4\xi\cos(2\theta),
\label{eq:r_sep}
\end{equation}
which corresponds to the polar equation of a Bernoulli lemniscate, separating trajectories confined to a single well from those that explore both wells. 

We want to determine the area enclosed by the separatrix.
The area element is $dA=dq dp = r dr d\theta$. For a closed curve  described by $r(\theta)$, the enclosed area is
\begin{equation}
A=\int dA = \int_{\theta_1}^{\theta_2 }\int_0^{r(\theta)}r dr d\theta = \int_{\theta_1}^{\theta_2 } \frac{r(\theta)^2}{2}  d\theta.
\end{equation}
Since $r^2$ in Eq.~\eqref{eq:r_sep} must be nonnegative, we need to ensure that $\cos(2\theta)\ge0$. This implies that one lobe of the Bernoulli lemniscate corresponds to the interval
\begin{equation}
-\frac{\pi}{2}\le2\theta\le\frac{\pi}{2} \Rightarrow -\frac{\pi}{4}\le\theta\le\frac{\pi}{4},
\end{equation}
and the other to 
\begin{equation}
\frac{3\pi}{2}\le2\theta\le\frac{5\pi}{2} \Rightarrow \frac{3\pi}{4}\le\theta\le\frac{5\pi}{4}.
\end{equation}
Thus there are two disconnected angular intervals.

Using Eq.~\eqref{eq:r_sep}
\begin{align}
A_{\rm lobe}
&=
\frac{1}{2}
\int_{-\pi/4}^{\pi/4}
4\xi\cos(2\theta)\,d\theta
=
2\xi.
\end{align}
Since the separatrix consists of two identical lobes, the total enclosed area is
\begin{equation}
A=2A_{\rm lobe}=4\xi.
\end{equation}

\section{Classical phase-space dynamics of the open KPO}
\label{app:open_kpo_classical}

In this section, we derive the classical dissipative flow used in the main text and analyze its stationary points and their stability. We also explain the coalescence of the attractors at the origin when $\kappa \rightarrow \kappa_c$.

\subsection{Classical dissipative flow}

 The Lindblad master equation is
\begin{equation}
    \frac{d\hat\rho}{dt}
    =
    \mathcal{L}\hat\rho
    =
    -i[\hat H,\hat\rho]
    + \kappa (1+n_{\rm th}) \mathcal{D}[\hat a]\hat\rho
    + \kappa n_{\rm th} \mathcal{D}[\hat a^\dagger]\hat\rho ,
    \label{Eq:Lindblad_app}
\end{equation}
where the Hamiltonian is in Eq.~(\ref{EqSM_Hq}) and 
\begin{equation}
\mathcal{D}[\hat O]\hat\rho
=
\hat O\hat\rho\hat O^\dagger
-
\frac{1}{2}
\left(
\hat O^\dagger \hat O \hat\rho
+
\hat\rho \hat O^\dagger \hat O
\right).
\end{equation}

For any operator $\hat O$, the equation of motion for its expectation value is
\begin{align}
\frac{d}{dt}\langle \hat O\rangle
&=
-i\langle[\hat O,\hat H]\rangle \nonumber \\
&+
\kappa(1+n_{\rm th})
\left\langle
\mathcal{D}^\dagger[\hat a]\hat O
\right\rangle
+
\kappa n_{\rm th}
\left\langle
\mathcal{D}^\dagger[\hat a^\dagger]\hat O
\right\rangle .
\end{align}
Applying this equation to $\hat O=\hat a$, we obtain
\begin{equation}
\frac{d}{dt}\langle\hat a\rangle
=
-2iK\langle\hat a^\dagger\hat a^2\rangle
+
2i\epsilon_2\langle\hat a^\dagger\rangle
-
\frac{\kappa}{2}\langle\hat a\rangle .
\label{EqSM_a}
\end{equation}

To derive Eq.(\ref{EqSM_a}) above, we use
\begin{equation}
[\hat a,\hat a^{\dagger 2}\hat a^2]
=
[\hat a,\hat a^{\dagger 2}]\hat a^2
=
2\hat a^\dagger \hat a^2,
\end{equation}
and
\begin{equation}
[\hat a,\hat a^{\dagger 2}+\hat a^2]
=
2\hat a^\dagger 
\end{equation}
in the Hamiltonian contribution $-i\langle[\hat a,\hat H]\rangle$. We also evaluate the dissipative contribution. For the loss term,
\begin{equation}
\mathcal{D}^\dagger[\hat a]\hat a
=
-\frac{1}{2}\hat a,
\end{equation}
while for the thermal excitation term,
\begin{equation}
\mathcal{D}^\dagger[\hat a^\dagger]\hat a
=
+\frac{1}{2}\hat a,
\end{equation}
so
\begin{align}
\frac{d}{dt}\langle\hat a\rangle_{\rm diss}
&=
\kappa(1+n_{\rm th})
\left(
-\frac{1}{2}\langle\hat a\rangle
\right)
+
\kappa n_{\rm th}
\left(
+\frac{1}{2}\langle\hat a\rangle
\right)
\nonumber\\
&=
-\frac{\kappa}{2}\langle\hat a\rangle .
\end{align}
One sees that the thermal occupation $n_{\rm th}$ cancels from the deterministic equation of motion for the field amplitude. It contributes to fluctuations and diffusion, but not to the classical drift of the mean field.

In the classical or mean-field limit, we replace
\begin{equation}
\langle\hat a\rangle\rightarrow \alpha,
\qquad
\langle\hat a^\dagger\rangle\rightarrow \alpha^*,
\qquad
\langle\hat a^\dagger\hat a^2\rangle
\rightarrow
|\alpha|^2\alpha .
\end{equation}
This yields the classical complex-amplitude equation
\begin{equation}
\dot\alpha
=
-2iK|\alpha|^2\alpha
+
2i\epsilon_2\alpha^*
-
\frac{\kappa}{2}\alpha .
\label{Eq:alpha_dot_app}
\end{equation}

We now introduce real phase-space variables $q$ and $p$ through
\begin{equation}
\alpha=\frac{q+ip}{\sqrt{2}},
\qquad
\alpha^*=\frac{q-ip}{\sqrt{2}}.
\end{equation}
Then
\begin{equation}
|\alpha|^2=\frac{q^2+p^2}{2},
\qquad
\dot\alpha=\frac{\dot q+i\dot p}{\sqrt{2}}.
\end{equation}
Substituting these expressions into Eq.~\eqref{Eq:alpha_dot_app}, we obtain
\[
\frac{\dot q+i\dot p}{\sqrt{2}}
=
-2iK
\left(
\frac{q^2+p^2}{2}
\right)
\frac{q+ip}{\sqrt{2}}
+
2i\epsilon_2
\frac{q-ip}{\sqrt{2}}
-
\frac{\kappa}{2}
\frac{q+ip}{\sqrt{2}} ,
\]
and, by multiplying by $\sqrt{2}$, we have 
\[
\dot q+i\dot p
=
-iK(q^2+p^2)(q+ip)
+
2i\epsilon_2(q-ip)
-
\frac{\kappa}{2}(q+ip).
\]
Separating real and imaginary parts, we find the classical dissipative flow of the open KPO,
\begin{align}
\dot q
&=
Kp(q^2+p^2)
+
2\epsilon_2 p
-
\frac{\kappa}{2}q,
\label{eq:app_flow_q}
\\
\dot p
&=
-Kq(q^2+p^2)
+
2\epsilon_2 q
-
\frac{\kappa}{2}p.
\label{eq:app_flow_p}
\end{align}

The equations of motion define a vector field. At every point in phase space, the vector field specifies the direction of the trajectory. The divergence measures the local expansion or contraction of this vector field.
The divergence of the flow in Eqs.(\ref{eq:app_flow_q})-(\ref{eq:app_flow_p}) is
\begin{align}
\nabla\cdot(\dot q,\dot p)
&=
\frac{\partial \dot q}{\partial q}
+
\frac{\partial \dot p}{\partial p}
=
\left(2Kpq-\frac{\kappa}{2}\right)
+
\left(-2Kpq-\frac{\kappa}{2}\right)
\nonumber\\
&=
-\kappa .
\end{align}
Thus, for $\kappa>0$, phase-space areas contract exponentially,
\begin{equation}
A(t)=A(0)e^{-\kappa t}.
\end{equation}
This is the classical signature of dissipation. In contrast, for the closed system, \(\kappa=0\), the phase-space flow is incompressible, as required by Liouville's theorem.

\subsection{Stationary points}
To determine the stationary points, it is convenient to use the polar coordinates in Eq.~(\ref{EqSM_polar}).
From Eqs.~\eqref{eq:app_flow_q}-\eqref{eq:app_flow_p}, we obtain
\begin{align}
\dot r
&=
r
\left(
2\epsilon_2\sin 2\theta
-
\frac{\kappa}{2}
\right),
\\
\dot\theta
&=
2\epsilon_2\cos 2\theta
-
Kr^2 .
\end{align}
Introducing
\begin{equation}
\kappa_c=4\epsilon_2,
\end{equation}
the nonzero stationary points satisfy
\begin{equation}
\sin 2\theta=\frac{\kappa}{\kappa_c},
\qquad
Kr^2=\frac{\kappa_c}{2}\cos 2\theta .
\end{equation}
They therefore exist only when
\begin{equation}
\kappa<\kappa_c.
\end{equation}
Notice that the Kerr nonlinearity $K$ does not play a role in the threshold for the existence of the nonzero attractors. The critical value $\kappa_c$ for their existence is set by the
competition between the amplitude of the parametric drive $\epsilon_2$ and the linear dissipation $\kappa$. The Kerr nonlinearity determines only the location of the nonzero attractors after they appear.

Defining
\begin{equation}
f=\kappa_c^2-\kappa^2,
\end{equation}
we have
\begin{equation}
\cos^2 (2\theta) = 1 - \sin^2 ( 2\theta) \Rightarrow \cos 2\theta=\frac{\sqrt f}{\kappa_c},
 \nonumber
\end{equation}
so
\begin{align}
  r &=\frac{ f^{1/4}}{\sqrt{2K}}, \nonumber \\
  \cos \theta &= \sqrt{\frac{\kappa_c + \sqrt{f}}{2 \kappa_c}} ,\nonumber \\
  \sin \theta &= \sqrt{\frac{\kappa_c - \sqrt{f}}{2 \kappa_c}} .
  \nonumber \\
\end{align}
This implies that the two nonzero fixed points can be written as
\begin{align}
x_1&=(q_1, \, p_1)
=
\left(
\sqrt{\frac{\sqrt f(\kappa_c+\sqrt f)}
{4K\kappa_c}} \,
\sqrt{\frac{\sqrt f(\kappa_c-\sqrt f)}
{4K\kappa_c}}
\right),
\nonumber\\
x_2&=(-q_1, \, -p_1).
\label{Eq:x1x2}
\end{align}
The third stationary point is the origin,
\begin{equation}
x_3=(0, \, 0).
\label{EqOrigin}
\end{equation}

\subsection{Stability of the stationary points}

We now analyze the stability of the stationary points. The Jacobian matrix of the flow is
\begin{equation}
J(q,p)
=
\begin{pmatrix}
\dfrac{\partial \dot q}{\partial q} &
\dfrac{\partial \dot q}{\partial p}
\\[8pt]
\dfrac{\partial \dot p}{\partial q} &
\dfrac{\partial \dot p}{\partial p}
\end{pmatrix},
\end{equation}
with
\begin{align}
\frac{\partial \dot q}{\partial q}
&=
2Kpq-\frac{\kappa}{2},
\\
\frac{\partial \dot q}{\partial p}
&=
2\epsilon_2+K(q^2+3p^2),
\\
\frac{\partial \dot p}{\partial q}
&=
2\epsilon_2-K(3q^2+p^2),
\\
\frac{\partial \dot p}{\partial p}
&=
-2Kpq-\frac{\kappa}{2}.
\end{align}

At the origin, the Jacobian becomes
\begin{equation}
J(0,0)
=
\begin{pmatrix}
-\kappa/2 & 2\epsilon_2
\\
2\epsilon_2 & -\kappa/2
\end{pmatrix}.
\end{equation}
Its eigenvalues are
\begin{equation}
\lambda_\pm^{(0)}
=
-\frac{\kappa}{2}
\pm
2\epsilon_2
=
-\frac{\kappa}{2}
\pm
\frac{\kappa_c}{2}.
\end{equation}
Thus, for \(\kappa<\kappa_c\), one eigenvalue is positive and one is negative, so the origin is a hyperbolic fixed point. For $\kappa>\kappa_c$, both eigenvalues are negative and the origin becomes a stable node. At $\kappa=\kappa_c$, one eigenvalue vanishes, signaling the bifurcation at which the two nonzero attractors merge with the origin.

At the nonzero fixed points \(x_1\) and \(x_2\), the eigenvalues of the Jacobian are
\begin{equation}
\lambda_\pm^{(1,2)}
=
-\frac{\kappa}{2}
\pm
\frac{1}{2}
\sqrt{5\kappa^2-4\kappa_c^2}.
\end{equation}
For all \(\kappa<\kappa_c\), the real parts of these eigenvalues are negative, so \(x_1\) and \(x_2\) are stable attractors. Their precise character depends on the sign of the discriminant. When
\begin{equation}
5\kappa^2-4\kappa_c^2<0 \Rightarrow \kappa<\frac{2}{\sqrt5}\kappa_c,
\end{equation}
the eigenvalues are complex with negative real part, and the fixed points are stable spiral attractors. When
\begin{equation}
\frac{2}{\sqrt5}\kappa_c<\kappa<\kappa_c,
\end{equation}
the eigenvalues are real and negative, and the fixed points are stable nodes.

This analysis shows that weak dissipation does not destroy the two-region organization inherited from the closed double-well KPO. Instead, the stable minima of the closed system are transformed into stable attractors of the dissipative flow, while the hyperbolic point separating the two regions persists. This remnant phase-space structure provides the classical mechanism underlying the pairwise organization of the spectrum of the steady-state Hamiltonian $H_{\rm ss}$.

%%%%%%%%%%%%
\section{Thermal interpretation of the spacing between paired eigenvalues}
\label{Appendix:harmonic-oscillator}

For $\xi \gg 1$ and $\kappa/K \ll \kappa_c/K$, Fig.~2(b) of the main text shows that the spacings between the neighboring pairs of
eigenvalues of $H_{\rm ss}$ become approximately constant and given by 
\begin{equation}
\eta_{m+1}-\eta_m
=
\ln\!\left(
\frac{n_{\rm th}+1}{n_{\rm th}}
\right).
\label{eq:thermalspacing}
\end{equation}
Here, $m$ labels successive local harmonic-oscillator
levels, each associated with a quasi-degenerate pair of eigenvalues of $H_{\rm ss}$. The approximation in Eq.~(\ref{eq:thermalspacing}) is particularly accurate for eigenstates far below the boundary separating the region of steady-state spectral kissing from nondegenerate eigenvalues. We now provide a physical explanation for its origin.

\subsection{Isolated KPO and harmonic oscillator}

In the isolated case with large squeezing amplitude $\xi=\epsilon_2/K$, the classical
phase space consists of two deep wells. Expanding the Hamiltonian around
either minimum yields an effective harmonic oscillator with frequency
\begin{equation}
\omega_{\rm loc}\sim 4\epsilon_2.
\end{equation}
This can be shown as follows.

We expand around either minimum in Eq.~(\ref{Eq:MinimaKPOisolated}),
\begin{equation}
q=q_0+\delta q,
\qquad
p=\delta p,
\end{equation}
and keep terms up to second order 
\begin{equation}
H_{\rm cl}
\simeq
H_{\rm cl}(q_0,0)
+
\frac{1}{2}
\left.
\frac{\partial^2 H_{\rm cl}}{\partial q^2}
\right|_{(q_0,0)} \!\! \!\! \!\!
(\delta q)^2
+
\frac{1}{2}
\left.
\frac{\partial^2 H_{\rm cl}}{\partial p^2}
\right|_{(q_0,0)} \!\! \!\! \!\!
(\delta p)^2 .
\end{equation}
The second derivatives are
\begin{align}
\frac{\partial^2 H_{\rm cl}}{\partial q^2}
&=
K(3q^2+p^2)-2\epsilon_2,\\
\frac{\partial^2 H_{\rm cl}}{\partial p^2}
&=
K(q^2+3p^2)+2\epsilon_2,
\end{align}
and evaluating them at $q_0$ and $p_0$ gives
\begin{equation}
H_{\rm cl}
\simeq
-\frac{(\epsilon_2)^2}{K}
+
2\epsilon_2\left[(\delta q)^2+(\delta p)^2\right],
\end{equation}
where $E_{\rm min} = -(\epsilon_2)^2/K$.

Using the canonical equations of motion for this Hamiltonian,
\begin{equation}
\dot{\delta q}
=
\frac{\partial H_{\rm cl}}{\partial \delta p}
=
4\epsilon_2\,\delta p,
\qquad
\dot{\delta p}
=
-\frac{\partial H_{\rm cl}}{\partial \delta q}
=
-4\epsilon_2\,\delta q,
\end{equation}
we obtain
\begin{equation}
\ddot{\delta q}
=
-16 (\epsilon_2)^2\,\delta q.
\end{equation}
Thus, for large $\xi$, the two wells are deep and well separated, and the dynamics near either minimum is approximately that of a harmonic oscillator with frequency $\omega_{\rm loc} \sim 4\epsilon_2$.

The quantum Hamiltonian corresponding to this scenario can be written as
\begin{equation}
H_{\rm loc}
=
E_{\rm min}
+
\omega_{\rm loc}
\, b^\dagger b ,
\end{equation}
where $b^\dagger$ and $b$ are the bosonic creation and annihilation
operators. 

\subsection{Weak coupling to the bath}

In the weak-dissipation
regime, $\kappa/K\ll\xi$, the attractors of the open system remain close to the minima. The Lindblad equation describing the local dynamics takes the form of a
harmonic oscillator coupled to a thermal bath,
\begin{equation}
\frac{d\rho}{dt}
=
-i[H_{\rm loc},\rho]
+
\kappa(n_{\rm th}+1)\,
{\cal D}[b]\rho
+
\kappa n_{\rm th}\,
{\cal D}[b^\dagger]\rho .
\label{eq:thermalME}
\end{equation}
Although the driven-dissipative KPO generally breaks global detailed balance due to continuous external pumping~\cite{Marthaler2006,Bachtold2022}, the leading-order local master equation in Eq.~\eqref{eq:thermalME} satisfies detailed balance with respect to $H_{\rm loc}$. Consequently, within this approximation, the local steady state corresponds to a thermal Gibbs state
\begin{equation}
\rho_{\rm th}
=
\frac{e^{-\beta H_{\rm loc}}}
     {\mathrm{Tr}\!\left(e^{-\beta H_{\rm loc}}\right)},
\label{eq:rho_th_gibbs}
\end{equation}
with inverse temperature $\beta$~\cite{Agarwal1971}.

Using the local Fock basis for a single well, 
\[
H_{\rm loc}|m\rangle
=
(E_{\rm min}+m\omega_{\rm loc})|m\rangle.
\]
Since the constant $E_{\rm min}$ cancels upon normalization of the thermal density matrix, we can omit it and write
\begin{align}
&e^{-\beta H_{\rm loc}}
=
\sum_{m=0}^{\infty}
e^{-m\beta\omega_{\rm loc}}
|m\rangle\langle m|, \nonumber \\
&\mathrm{Tr}\!\left(e^{-\beta H_{\rm loc}}\right)
=
\sum_{m=0}^{\infty}
e^{-m\beta\omega_{\rm loc}}
=
\frac{1}{1-e^{-\beta\omega_{\rm loc}}}.
\end{align}

Therefore,
\begin{align}
\rho_{\rm th}
&=
\left(1-e^{-\beta\omega_{\rm loc}}\right) \sum_{m=0}^{\infty}
e^{-m\beta\omega_{\rm loc}}   
|m\rangle\langle m| \nonumber \\
&= 
\sum_{m=0}^{\infty}
p_m |m\rangle\langle m|,
\label{eq:rho_expansion}
\end{align}
where $p_m$ are the eigenvalues of the steady state density matrix.

The thermal occupation number of the local oscillator is
\begin{equation}
n_{\rm th}
=
\frac{1}{e^{\beta\omega_{\rm loc}}-1},
\end{equation}
which implies that 
\begin{equation}
e^{-\beta\omega_{\rm loc}}
=
\frac{n_{\rm th}}
     {n_{\rm th}+1}.
     \label{eq:exp_omega}
\end{equation}
From Eq.~(\ref{eq:rho_expansion}), Eq.~(\ref{eq:exp_omega}), and normalizing the total probability, we obtain
\begin{equation}
p_m
=
\frac{1}{2}\frac{1}{n_{\rm th}+1}
\left(
\frac{n_{\rm th}}
     {n_{\rm th}+1}
\right)^m.
\label{eq:pm}
\end{equation}

Since the steady-state Hamiltonian is defined through
$\rho_{\rm ss}=e^{-H_{\rm ss}}$,
the corresponding eigenvalues of $H_{\rm ss}$ are
\begin{align}
\eta_m
&=
\ln 2+
\ln(n_{\rm th}+1)
+
m\ln\!\left(
\frac{n_{\rm th}+1}{n_{\rm th}}
\right).
\end{align}
Consequently, the spacing between neighboring pairs of quasi-degenerate eigenvalues satisfies
Eq.~(\ref{eq:thermalspacing}).
The spectrum of $H_{\rm ss}$ can be viewed as two nearly identical copies of the thermal spectrum, leading to quasi-degenerate pairs of eigenvalues separated according to Eq.~(\ref{eq:thermalspacing}). 
%%%%%%%

%%%%%%%%%%%%%%%%%%%%%%%%%%%%%%%%%%%%%%%%%%%%%%%%%%%%
%%%%%%%%% REFERENCES %%%%%%%%%%%%%%%%%%%%%%%%%%%%%%%
%%%%%%%%%%%%%%%%%%%%%%%%%%%%%%%%%%%%%%%%%%%%%%%%%%%%

\bibliography{biblio2025}

%%%%%%%%%%%%%%%%%%%%%%%%%%%%%%%%%%%%%%%%%%%%%%%%%%%%
%%%%%%%%%%%%%%%%%%%%%%%%%%%%%%%%%%%%%%%%%%%%%%%%%%%%
\end{document}